\documentclass[fleqn,twoside]{article}
\usepackage{espcrc2,epsfig}

\usepackage{graphicx}
\usepackage[figuresright]{rotating}

\newcommand{\AmS}{{\protect\the\textfont2
  A\kern-.1667em\lower.5ex\hbox{M}\kern-.125emS}}

\title{
{\vspace{-1.4em} \parbox{\hsize}{\hbox to \hsize
{\hss \normalsize HU-EP-02/39}
\parbox{\hsize}{\hbox to \hsize
{\hss \normalsize LU-ITP 2002/020}
\parbox{\hsize}{\hbox to \hsize
{\hss \normalsize Edinburgh 2002/10}
\parbox{\hsize}{\hbox to \hsize
{\hss 
\vspace{11pt}
\normalsize LTH 559}}}}}}\\
Finite Size Effects in Nucleon Masses in Dynamical QCD
\vspace{5pt}
\thanks{presented by A. Ali Khan}}

\author{A. Ali Khan\address[HU]{Institut f\"ur Physik, Humboldt-Universit\"at 
zu Berlin,  10115 Berlin, Germany}, 
T.~Bakeyev\address[DU]{Joint Institute for Nuclear Research, 141980 Dubna,
Russia},
M. G\"ockeler\address{Institut f\"ur Theoretische Physik, Universit\"at 
Leipzig, 04109 Leipzig, Germany}\address[RU]{Institut f\"ur Theoretische 
Physik, Universit\"at Regensburg, 93040 Regensburg, Germany}, 
R. Horsley\address[EU]{School of Physics, The University of 
Edinburgh, Edinburgh EH9 3JZ, UK}, 
A. C. Irving\address[LI]{Theoretical Physics Division, Department of 
Mathematical Sciences, University of Liverpool, 
Liverpool L69 3BX, UK},
D. Pleiter\address[NIC]{John von Neumann-Institut f\"ur Computing NIC, 
15738 Zeuthen, Germany}, 
P. Rakow\addressmark[RU]\addressmark[LI], \\
G. Schierholz\addressmark[NIC]\address[DESY]{Deutsches 
Elektronen-Synchrotron  DESY, 22603 Hamburg, Germany}, 
and H. St\"uben\address{Konrad-Zuse-Zentrum f\"ur Informationstechnik
Berlin, 14195 Berlin, Germany}
(QCDSF and UKQCD Collaborations)
}
       
\begin{document}

\begin{abstract}
For lattice calculations with light dynamical quarks, finite size effects
have become an important aspect. We study finite size effects in nucleon 
masses on $N_f=2$ dynamical lattices of $1-2$ fm. Predictions for the finite 
size effects are obtained in one-loop chiral perturbation theory.

\vspace{1pc}
\end{abstract}

\maketitle

\section{INTRODUCTION}
For a lattice calculation of physical quantities such as hadron masses it
is desirable to have a theoretical understanding of finite size effects and
to be able to extrapolate to infinite volume. On lattices used in typical 
quenched calculations, finite size effects of hadron 
masses seem to be small, but they may be considerable in the dynamical 
case~\cite{irving02}. A source
of finite size effects is a virtual pion going around the boundary of the
lattice before it is absorbed again. At sufficiently small pion masses and
large volumes this effect can be described by chiral perturbation theory,
the low-energy effective theory of nucleons and pions, in a finite box.
%For sufficiently large volumes, one expects hadronic physics to be described 
%in a sensible manner with the QCD degrees of freedom at low energy, pions and 
%baryons. In this context, chiral perturbation theory $\chi PT$ is 
%considered to be useful.
Here, we discuss finite size effects on nucleon masses generated by the
QCDSF and UKQCD collaborations. 
\section{THE SIMULATION}
The simulations were done using the pla\-quette gauge action and $N_f=2$
dynamical non-perturbatively $O(a)$-improved Wilson fermions.
We consider lattices of size $\sim 1$ fm,  $\sim 1.6$ fm, and  $\sim 2$ fm. 
Pion  masses are in the range of about $0.6 - 1$ GeV. 
The lattice spacing is $a \approx 0.1$ fm. We use $r_0=0.5$ fm to set the 
scale. Valence and sea quark masses are taken to be equal.

\begin{figure}[htb]
\begin{center}
%\vspace{-2pt}
\epsfysize=6.7cm \epsfbox{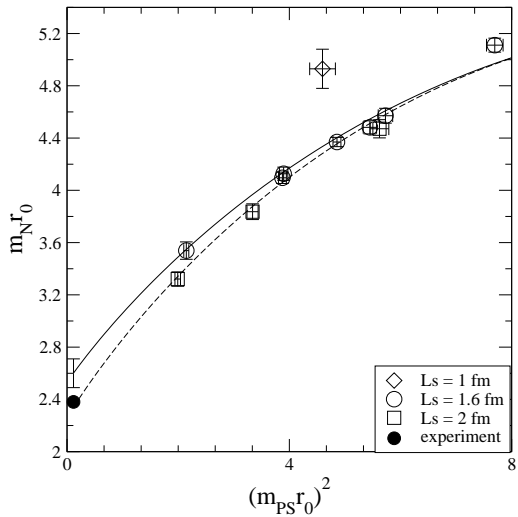}
\end{center}
%\vspace{-20pt}
\caption{Nucleon mass as a function of the pion mass with extrapolation to the
physical value according to Eq.~(\protect\ref{eq:extrapb}). The solid line
denotes a fit to the $1.6$ fm lattices. The dashed line corresponds to the
2 fm lattices, without error bar on the corresponding 
chiral extrapolation, since there are only three points. The 
heaviest point at $1.6$ fm is not included in the fit.
}
%\vspace{-5pt}
\label{fig:dta}
\end{figure}

The lattice data for the nucleon mass are plotted in Fig.~\ref{fig:dta}.
We see significant finite size effects. We expect $O(a^2)$ effects to be 
small. Indeed, our two mass points at $r_0 m_{PS} \simeq 2$ on the 1.6 fm 
lattice differ by less than a percent, whereas $a^2$ differs by $O(20)\%$. 

Chiral perturbation theory in the infinite volume predicts
\begin{eqnarray}
m_Nr_0 & = & a + b (m_{PS}r_0)^2 + c 
(m_{PS}r_0)^3, \label{eq:extrapb}
\end{eqnarray}
where the cubic term reflects the leading non-analytic (LNA) 
behaviour of the nucleon self-energy.
The chiral extrapolation is performed separately for different lattice sizes
using Eq.~(\ref{eq:extrapb}). 
%Fitting all lattices with $L_s \geq 1.6$ fm 
%together, one would find $\chi^2/d.o.f. = 1$.  
At 1.6 fm, the chiral extrapolation  gives a value that is 8\% or
$2\sigma$ higher than experiment, at 2 fm no discrepancy is found.
%The coefficient $c$ is determined with an accuracy of $\sim 30\%$ on the 
%$1.6$ fm lattices. 
The coefficient $c$ disagrees significantly with 
the infinite volume LNA prediction. 
%If $c$ in Eq.~\ref{eq:extrapb} is fixed to zero, an extrapolation with
%$\chi^2/d.o.f. < 1$ is still possible at $1.6$ fm, the discrepancy 
%with experiment being $30\%$.
% Comparing two results at the same  pion mass of
% $r_0m_{PS} \simeq 2$,  where $a$ differs by $8\%$,
% we find a  small 
% difference in $r_0m_N$ $< 1\%$, or $1/2$ of the statistical error. 
% \begin{table}[hbt]
% \caption{$m_Nr_0$ extrapolated to the experimental value.
%  $r_0(\mathrm{experimental}) = 0.5$ fm.}
% \label{tab:table}
% \begin{tabular}{|ccc|}
% \hline
% approx.          & extrapolated &  discrepancy  \\
% lattice size[fm] & value  &  to experiment \\
% \hline           
% \multicolumn{3}{|c|}{Eq.~\protect\ref{eq:extrapa}} \\
% \hline
% 1.6              & $3.05(5)$ & $28\%,5.5\sigma$ \\
% 2                & $2.82(9)$ &  $18\%,2\sigma$         \\
% \hline
% \multicolumn{3}{|c|}{Eq.~\protect\ref{eq:extrapb}} \\
% \hline
% 1.6               & $2.83(10)$ & $10\%,4.5\sigma$ \\
% 2                 &    $2.47$ & $4\%$ \\
% \hline
% \end{tabular}
% \end{table}
%
\section{THEORETICAL PREDICTIONS}
\begin{figure}[t]
%\vspace{-3pt}
\centerline{\epsfysize=1.1cm \epsfbox{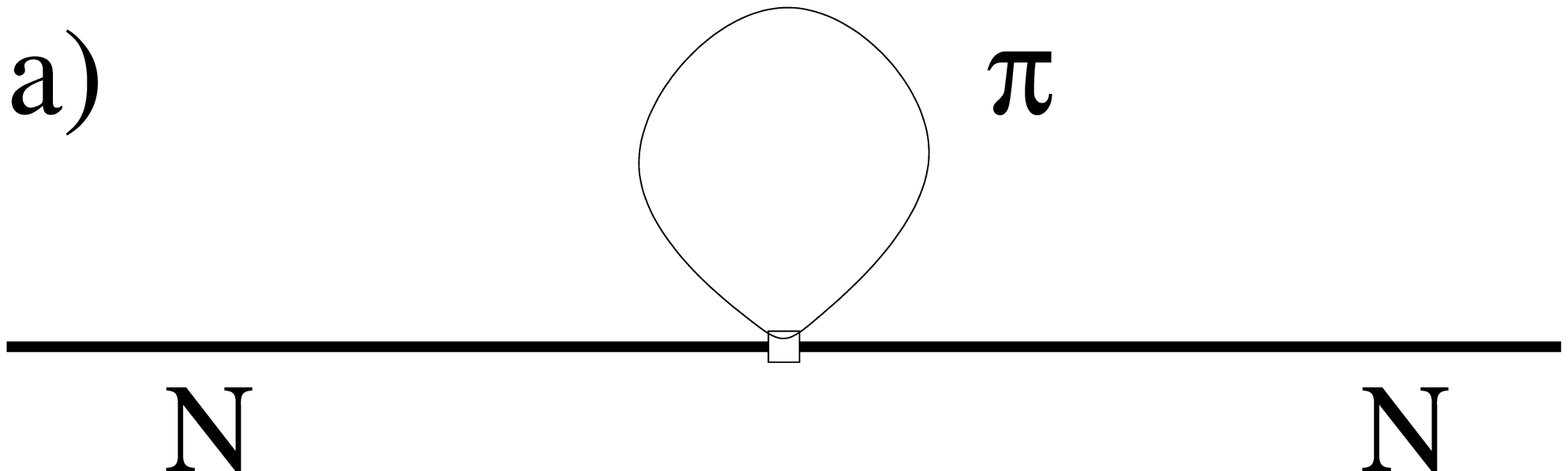}
           \epsfysize=1.1cm \epsfbox{empty.eps}}
\vspace{15pt}
\centerline{\epsfysize=1.1cm \epsfbox{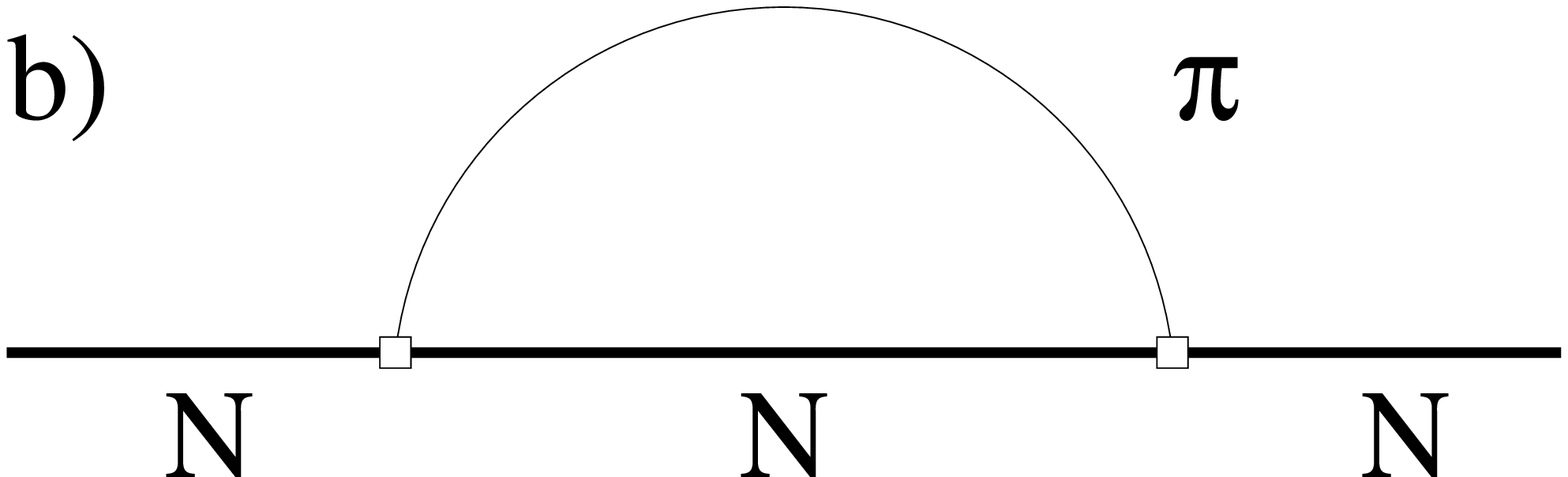}
           \epsfysize=1.1cm \epsfbox{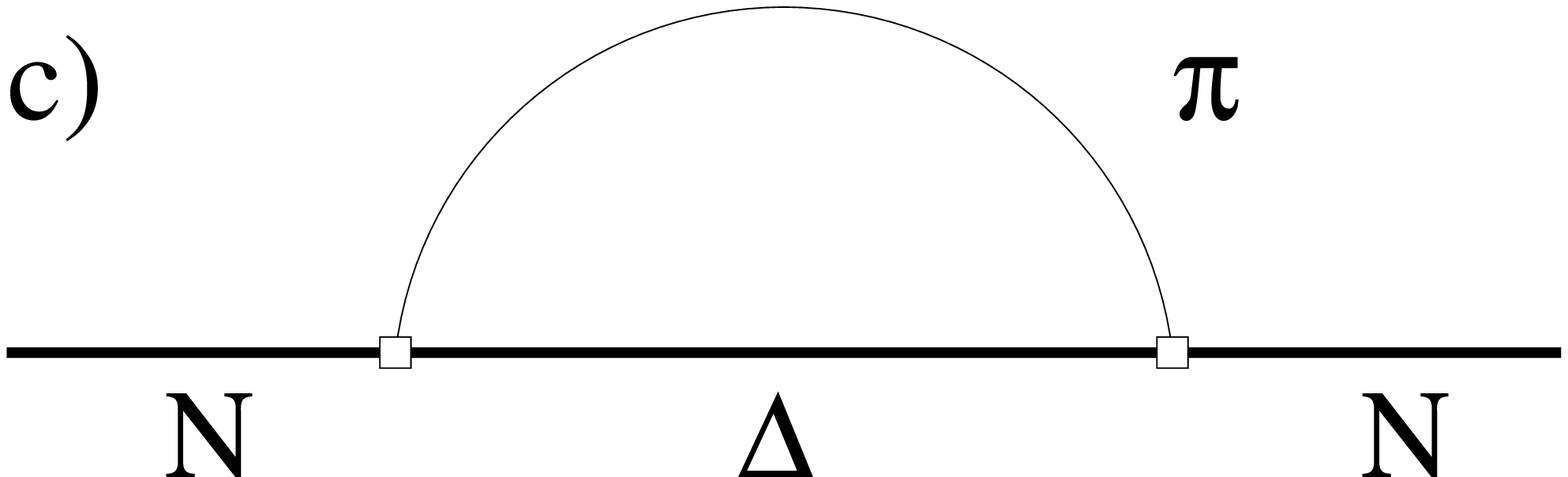}}
\vspace{-5pt}
\caption{Self-energy diagrams relevant for the calculation at one loop.}
\label{fig:diag}
\end{figure}

The finite size effects are calculated from the difference of the
nucleon self-energy $\Sigma$ in finite and infinite spatial volume,
\begin{equation}
\delta m_N = \Sigma(L_s)-\Sigma(\infty). \label{eq:FSE}
\end{equation}
In the perturbative calculation, the 
$x_4$ direction is taken to be of infinite extent. 
We calculate the self-energy at one loop in Heavy Baryon $\chi PT$ 
according to the lattice prescription given in Ref.~\cite{le01} 
adapted to the case of 2-flavours.
Additionally, we include $\Delta$ intermediate states. For this we 
discretise the 
$N\Delta\pi$ interaction Lagrangian given in Ref.~\cite{hem02}. The numerical
values for $f_\pi$, $g_A = {\cal D} + {\cal F}$, and the $N\Delta\pi$
coupling $c_A$ are also taken from this reference. 
The relevant Feynman diagrams are given in Fig.~\ref{fig:diag}. 

\begin{figure}[htb]
%\vspace{-9pt}
\epsfysize=6.7cm \epsfbox{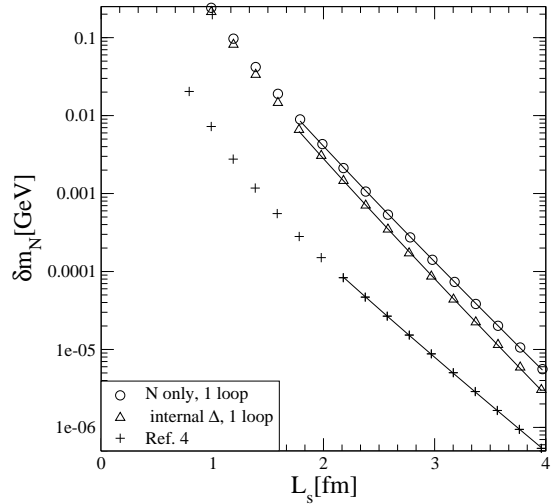}
%\vspace{-15pt}
\caption{$L_s$ dependence of $\delta m_N$ for $m_{PS} = 578$ MeV. Circles 
denote the 
sum of diagrams a) and b) of Fig.~\protect\ref{fig:diag}, triangles the
contribution of diagram c) of Fig.~\protect\ref{fig:diag}, and plusses the 
result of Ref.~\protect\cite{lusch83}. Fits to $\delta m_N=(c_0/L_s)
 \exp(-c_1L_s)$ are drawn as solid lines.
}
%\vspace{-5pt}
\label{fig:comp}
\end{figure}

To regularize the lattice integrals we vary the cut-off $\Lambda$ between 
$\pi/a = 6$ GeV and $\infty$. The difference between using  $\Lambda = 6$ 
GeV and the continuum limit is around $10\%$ for $m_{PS}\simeq 600$ MeV.  
The values presented here are for $\Lambda \approx 30$ GeV, which  
practically amounts to infinite cut-off.
Results for the smallest pion mass used in the simulation at $1.6$ fm 
are given  in Fig.~\ref{fig:comp}.
For $L_s \geq 1.6$ fm, the $L_s$ dependence can be approximated by an 
exponential decay,
\begin{equation}
\delta m_N = (c_0/L_s) \exp(-c_1 L_s), \label{eq:exp}
\end{equation}
where the fit with Eq.~(\ref{eq:exp}) describes the $\chi PT$ result with an 
accuracy of several percent.

% The range of momentum scales where $\chi PT$ is expected to be valid is 
%$\leq 1
% \,\mathrm{GeV}$. We would like to note the main contribution to the
% self-energy c
The finite size effect of the nucleon mass has been estimated previously,
relating the mass shift to the $\pi N$ scattering amplitude~\cite{lusch83}. 
A comparison with our one-loop result is given in Fig.~\ref{fig:comp}.
%A discussion of possible explanations for the large difference between
%L\"uscher's result and ours is beyond the scope of this paper.

\begin{figure}[htb]
%\vspace{-9pt}
\epsfysize=6.7cm 
\epsfbox{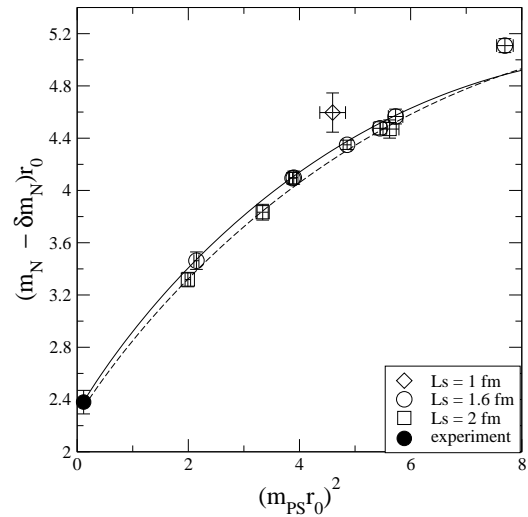}
%\vspace{-15pt}
\caption{Nucleon masses minus the finite size effect calculated in
one-loop $\chi PT$. As in Fig.~\protect\ref{fig:dta}, lines denote fits 
according to Eq.~(\protect\ref{eq:extrapb}). }
\label{fig:extrap}
\end{figure}

In Fig.~\ref{fig:extrap}, we show $m_N - \delta m_N$. We see that
the masses on the smaller volumes are brought down towards a
universal curve. In particular, 50\% of the finite size corrections
at $L_s = 1$ fm, and 60\% at $L_s = 1.6$ fm, are accounted for by chiral
perturbation theory. We consider this a remarkable result, considering
that the pion mass is relatively heavy.
\section{CONCLUSIONS AND DISCUSSION}
We find indications for finite size effects in the nucleon mass 
on lattices of $1.6-2$ fm size.
We calculated the finite volume corrections in lattice-regularised
$\chi PT$ at one loop. At $L_s \geq 1.6$ fm, the correction is of the 
order of  a few percent for the pion masses considered. $\chi PT$ describes 
about $50-60\%$ of the finite size effects of the data. The mass dependence is 
exponential rather than power-like even for intermediate lattices and pion
masses. 
Higher order corrections to the one-loop calculation may be important
at present pion masses.
\section*{ACKNOWLEDGEMENTS} 
We acknowledge useful discussions with Th.~Hemmert.
The numerical calculations were performed on the Hitachi {SR8000} at
  LRZ (Munich), the APEmille at NIC (Zeuthen), the Cray {T3E} at EPCC
  (Edinburgh) and NIC (J\"{u}lich).  We wish to thank all institutions
  for their support. This work has been supported in part by the European
Community's Human Potential Program under contract HPRN-CT-2000-00145,
Hadrons/Lattice QCD.

\end{document}